\documentclass[conference]{IEEEtran}

\usepackage[usenames]{color}
\usepackage[bookmarksnumbered=true]{hyperref}
\usepackage{url}
\definecolor{cite_color}{RGB}{0, 51, 153}
\definecolor{link_color}{RGB}{0, 120,0}
\definecolor{url_color}{RGB}{153, 102,  0}
\hypersetup{
 colorlinks,
 citecolor=cite_color,
 linkcolor=link_color,
 urlcolor=url_color}

\usepackage{amsmath}
\usepackage{amssymb}
\usepackage{amsthm, mathtools}
\newtheorem{theorem}{Theorem}

\usepackage{graphicx}
\usepackage{subfig}
\usepackage{multirow}
\usepackage{booktabs}

\usepackage[vlined, ruled, linesnumbered]{algorithm2e}

\SetCommentSty{mycommfont}

\usepackage{mathrsfs}
\usepackage{wrapfig}
\usepackage{color}
\newcommand{\MAXCUT}{\textsc{MaxCut}}
\usepackage{enumerate}

\def\C{{\cal C}}

\newcommand{\set}[1]{\{#1\}}

\newcommand{\sym}{$\bullet$\;}

\ifCLASSINFOpdf
\else
\fi
\begin{document}
%
\title{Information Content of Greedy \\\MAXCUT\  Algorithms}
\title{Greedy \MAXCUT\  Algorithms \\ and their Information Content}

\author{\IEEEauthorblockN{Yatao Bian, Alexey Gronskiy and Joachim
    M. Buhmann}
  \IEEEauthorblockA{Department of Computer Science, ETH Zurich\\
   \texttt{ \{ybian, alexeygr, jbuhmann\}@inf.ethz.ch}}
}


\maketitle

\begin{abstract}
  \MAXCUT\ defines a classical NP-hard problem for graph partitioning
  and it serves as a typical case of the symmetric non-monotone
  Unconstrained Submodular Maximization (USM) problem.  Applications
  of \MAXCUT\ are abundant in machine learning, computer vision and
  statistical physics. Greedy algorithms to approximately solve
  \MAXCUT\ rely on greedy vertex labelling or on an edge contraction
  strategy. These algorithms have been studied by measuring their
  approximation ratios in the worst case setting but very little is
  known to characterize their robustness to noise contaminations of
  the input data in the average case. Adapting the framework of
  Approximation Set Coding, we present a method to exactly measure the
  cardinality of the algorithmic approximation sets of five greedy
  \MAXCUT\ algorithms. Their information contents are explored for
  graph instances generated by two different noise models: the edge
  reversal model and Gaussian edge weights model. The results provide
  insights into the robustness of different greedy heuristics and
  techniques for \MAXCUT, which can be used for algorithm design of
  general USM problems.
\end{abstract}

%
\IEEEpeerreviewmaketitle

\section{Introduction}

Algorithms are mostly analyzed by measuring their runtime and memory
consumption for the worst possible input instance. In many application
scenarios, algorithms are also selected according to their
``robustness'' to noise perturbations of the input instance and their
insensitivity to randomization during algorithm execution. How should
this ``robustness'' property be measured? Machine learning requires
that algorithms with random variables as input generalize over these
fluctuations. The algorithmic answer has to be stable w.r.t. this
uncertainty in the input instance. Approximation Set Coding (ASC)
quantifies the impact of input randomness on the solution space of an
algorithm by measuring the attainable resolution for the algorithm's
output. We employ this framework in an exemplary way by estimating the
robustness of \MAXCUT\ algorithms to specific input
instances. Thereby, we effectively perform an average case analysis of
the generalization properties of \MAXCUT\ algorithms.

\subsection{\MAXCUT\  and Unconstrained Submodular Maximization}

Given an undirected graph $G=(V, E)$ with vertex set $V=\{v_1,
v_2, \cdots, v_n\}$ and edge set $E$ with nonnegative weights
$w_{ij}, \forall (i,j)\in E$, the \MAXCUT\  problem aims to find a
partition of vertices into two disjoint subsets $S_1$ and $S_2$, such
that the cut value 
$cut(S_1, S_2):=\sum_{i\in S_1} \sum_{j\in S_2} w_{ij}$ is maximized.
\MAXCUT\ is emlpoyed in various applications, such as in
semisupervised learning (\cite{Wang:2013:SLU:2502581.2502605}), in
social network (\cite{agrawal2003mining}), in statistical physics and
in circuit layout design (\cite{barahona1988application}).  \MAXCUT\
is considered to be a typical case of the USM problem because its
objective can be formulated as a set function: $f(S):=cut(S,
V\backslash S)
, S\subseteq V$, which is submodular, nonmonotone, and
symmetric ($f(S) = f(V\backslash S)$).  Beside
\MAXCUT, USM captures many practical problems such as
\textsc{MaxDiCut} (\cite{halperin2001combinatorial}), variants of
\textsc{MaxSat} and the maximum facility location problem
(\cite{ageev19990,cornuejols1977uncapacitated}).

\subsection{Greedy Heuristics and Techniques}

The five algorithms investigated here
(as summarized in Table \ref{tab-alg-summarization}) belong to two
greedy \textit{heuristics}: double greedy and backward
greedy. The \textit{double} greedy algorithms
 exploit the symmetric property of USM, and conducts
classical forward greedy and backward greedy simultaneously: it works
on two solutions initialized as $\emptyset$ and the ground set $V$,
respectively, then processes the
%
elements (vertices for \MAXCUT\ problem) one at a time, for which it
determines whether it should be added to the first solution or removed
from the second solution.  The \textit{backward} greedy algorithm removes the
smallest weighted edge in each step.  The difference of the four
double greedy algorithms lies in the greedy \textit{techniques} they
use: sorting, randomization and the way to initialize the first two
vertices.


\subsection{Approximation Set Coding for Algorithm Analysis}

In analogy to Shannon's theory of communication, the ASC framework
(\cite{DBLP:dblp_conf/isit/Buhmann10},
\cite{JB:mcpr:2011}, \cite{buhmann2013simbad})
determines distinguishable sets of solutions and, thereby, provides a
general principle to conduct model validation
(\cite{DBLP:dblp_journals/jmlr/ChehreghaniBB12},
\cite{zhousparse}). As an
algorithmic variant of the ASC framework, \cite{busse2012information,
  informative-mst} defines the \textit{algorithmic $t$-approximation
  set} of an algorithm $\mathscr{A}$ at step $t$ as the set of
feasible solutions after $t$ steps, $C_t^{\mathscr{A}}(G):= A_t(G)$, where $A_t(G)$ is the solution set which are
still considered as viable by $\mathscr{A}$ after $t$ computational
steps.

%

ASC utilizes the \textit{two instance-scenario} to investigate the information
content of greedy \MAXCUT\ algorithms. Since we investigate the
average case behavior of algorithms, we have to specify the
probability distribution of the input instances. We
generate graph instances in a two step process.  
First,  generate a
``master graph'' $G$, e.g., a complete graph with Gaussian distributed
weights.
In a second step, we generate two input graphs
$G^\prime,\;G^{\prime\prime}$ by independently applying a noise
process to edge weights of the master graph $G$. 




The algorithmic analogy of \textit{information content}
(\cite{DBLP:dblp_conf/isit/Buhmann10}), i.e. algorithmic information
content $I^{\mathscr{A}}$, is computed as the maximum stepwise
information $I_t^{\mathscr{A}}$:
\begin{equation}\label{eq:ic}
  I^{\mathscr{A}} := 
\max_t  I_t^{\mathscr{A}} = 
  \max_t  \mathbb{E} \Bigl[ 
  \log \Bigl( \frac{ |\C|\; \Delta_t^{\mathscr{A}}(G',G'')}{|C_t^{\mathscr{A}}(G')| |C_t^{\mathscr{A}}(G'')|} \Bigr) 
  \Bigr]   
\end{equation}
The expectation is taken w.r.t. ($G'$, $G''$);
$\Delta_t^{\mathscr{A}}(G',G'') := |C_t^{\mathscr{A}}(G') \cap
C_t^{\mathscr{A}}(G'')|$ denotes the intersection of approximation
sets, and $\C$ is the solution space, i.e., all possible cuts.
The information content $ I^{\mathscr{A}}_t$ measures how much
information is extracted by algorithm ${\mathscr{A}}$ at iteration $t$
from the input distribution that is relevant to the output
distribution.

\begin{table}
\begin{center}
\normalsize
\caption{Summary of Greedy \MAXCUT\   Algorithms}
\label{tab-alg-summarization}
\begin{tabular}{|c|c|c|c|c|c|}
\hline
\multirow{2}{*}{Name}  & Greedy     & \multicolumn{3}{c|}{Techniques}  \\
   \cline{3-5}
     & Heuristic &  Sort. & Rand. & Init. Vertices \\
  \hline
  \hline
  D2Greedy  & \multirow{4}{*}{Double} &  &  &\\
  \cline{1-1}  \cline{3-5}
  RDGreedy  &  &  & $\checkmark$  &\\
  \cline{1-1}  \cline{3-5}
  SG & &  &  & $\checkmark$\\
  \cline{1-1}  \cline{3-5}
  SG3  &  &  $\checkmark$  &   & $\checkmark$\\
  \hline
  EC & Backward & $\checkmark$  &  &\\
  \hline
\end{tabular}
\end{center}
\end{table}

\section{Greedy \MAXCUT\  Algorithms}

We investigate five greedy algorithms (Table
\ref{tab-alg-summarization}) for \MAXCUT. According to the type of
greedy heuristic, they can be
divided into two categories: I)~\textit{Double Greedy}: SG, SG3,
D2Greedy, RDGreedy; II)~\textit{Backward Greedy}: Edge Contraction.
%
%
%
%
%
Besides the type of greedy heuristic, the difference between the
algorithms are mainly in three techniques: \textit{sorting} the
candidate elements, \textit{randomization} and the way
\textit{initializing the first two vertices}. In the following, we
briefly introduce one typical algorithm in each category and we
present the others by showing the difference (details are in the
Supplement
\ref{sup:alg} because of space limit).

\subsection{Double Greedy Algorithms}

D2Greedy (Alg. \ref{alg:d-usm-4-max-cut}) is the
{\textbf{D}}eterministic double greedy, RDGreedy is the
\textbf{R}andomized double greedy,
they were proposed by \cite{buchbinder2012tight} to solve the general
USM problem with ${1/3}$ and ${1/2}$ worst-case approximation
guarantee, respectively.  They use the same double greedy heuristic as
SG (\cite{sahni1976p}) and SG3 (variant of SG), which are classical
greedy \MAXCUT\ algorithms.  
%
We prove in Supplement \ref{app:equivalence-SG-D2Greedy} that, for
\MAXCUT, SG and D2Greedy use equivalent labelling criteria except for
initializing the first two vertices.


\begin{algorithm}
  \caption{D2Greedy
    (\cite{buchbinder2012tight})}\label{alg:d-usm-4-max-cut}
\begin{small}
\KwIn{Complete graph $G=(V, E)$ with nonnegative edges}
\KwOut{A disjoint cut  and the cut value}
{$S^0 :=\emptyset$, $T^0 := V$}\;
\For{$i=1$ to $n$}{
    {$a^i := f(S^{i-1} \cup \{v_i\})-f(S^{i-1})$}\;
    {$b^i := f(T^{i-1} \backslash \{v_i\})-f(T^{i-1})$}\;
    \If{$a^i\geq b^i$}
        {$S^i :=S^{i-1} \cup \set{v_i}$, $T^{i}:=T^{i-1}$ \tcp*[r]{expand $S$}} 
    \Else{$S^i :=S^{i-1}$, $T^{i}:=T^{i-1} \backslash \set {v_i}$ \tcp*[r]{shrink $T$}}
}
\KwRet{$S^n$, $V \backslash S^n$, and $cut(S^n, V \backslash S^n)$} 
\end{small}
\end{algorithm}

As shown in Alg. \ref{alg:d-usm-4-max-cut}, D2Greedy maintains two
solution sets: $S$ initialized as $\emptyset$, $T$ initialized as the
ground set $V$. It labels all the vertices one by one: for vertex
$v_i$, it computes the objective gain of adding $v_i$ to $S$ and the
gain of removing $v_i$ from $T$, then labels $v_i$ to have higher
objective gain.

SG and D2Greedy differ in the initialization of 
the first two vertices: SG picks first of all the maximum weighted
edge and distributes its two vertices to the two active subsets. Compared
to D2Greedy, the RDGreedy uses randomization technique when labelling
each vertex: it labels each vertex with probability proportional to
the objective gain. Compared to SG, SG3 sorts the unlabelled vertices
according to a certain score function (which is proportional to the
possible objective gains), and selects the vertex with the maximum
score to be the next one to be labelled.

\subsection{Edge Contraction (EC)}
\label{sec:ec}

EC (\cite{kahruman2007greedy}, Alg. \ref{alg:edge-contraction})
contracts the smallest edge in each step. The two vertices of this
contracted edge become one ``super'' vertex, and the weight of an edge
connecting this super vertex to any other vertex is assigned as the
sum of weights of the original two edges.  EC belongs to the backward
greedy in the sense that it tries to remove the least expensive edge from
the cut set in each step.  We can easily derive a heuristic for the
{\textsc{Max-k-Cut}} problem by using $n-k$ steps instead of $n-2$
steps.

\begin{algorithm}
\caption{Edge Contraction (EC) (\cite{kahruman2007greedy})}\label{alg:edge-contraction}
\begin{small}
\KwIn{ Complete graph $G=(V, E)$ with nonnegative edge}
\KwOut{A disjoint cut $S_1, S_2$ and cut value $cut(S_1, S_2)$}
\For{$i=1:n$}
    {$ContractionList(i):= \{i\}$\;}
\For{$i=1:n-2$}{
    {Find a minimum weight edge $(x, y)$ in $G$}\; 
    {$v:=contract(x,y)$, $V:=V\cup\{v\}\backslash\{x,y\}$ \tcp*[r]{contract}}   
    \For{$j\in V\backslash \{v\}$}
        {$w_{vj}:= w_{xj}+w_{yj}$\;}
    {$ContractionList(v) := ContractionList(x) \cup ContractionList(y)$}\;
}
{Denote by $x$ and $y$ the only 2 vertices in $V$}\;
\Return{$S_1:=ContractionList(x)$, 
$S_2:=ContractionList(y)$,
$cut(S_1, S_2):=w_{xy}$}
\end{small}
\end{algorithm}

%
%


\section{Counting Solutions in Approximation Sets}
\label{sec:counting}

To compute the information content according to Eq. \ref{eq:ic}, we
need to exactly compute the cardinalities of four different solution
sets. For \MAXCUT{} problem, the solution space has the cardinality
$|\C| = 2^{n-1} - 1$. In the following we will present guaranteed
methods for \textit{exact} counting $|C_t^{\mathscr{A}}(G')|,
|C_t^{\mathscr{A}}(G'')|$ and $\Delta_t^{\mathscr{A}}(G',G'')$
(sub-/superscripts omitted for notational clarity). 

%
%
%
%

\subsection{Counting Methods for Double Greedy Algorithms}
\label{sec:counting-sg3}

The counting methods for the double greedy algorithms are similar, so
we only discuss the method for SG3 here; details about other methods
and the corresponding proofs are in the Supplement
\ref{complement:counting} and \ref{app:proof-SG3}, respectively.

For the SG3 (Alg. \ref{alg:sg3}, see Supplement), after step $t$ ($t =
1, \cdots, n-1$) there are $k = n-t-1$ unlabelled vertices, and it is
clear that 
$|C(G')|=|C(G'')|=2^{k}$.

To count the intersection set $\Delta (G',G'')$, assume the solution
set pair of $G'$ is $(S_1', S_2')$, the solution set pair of $G''$ is
$(S_1'', S_2'')$, so the unlabelled vertex sets are $T'=V\backslash
\{S_1' \cup S_2'\}$, $T''=V\backslash \{S_1'' \cup S_2''\}$,
respectively.  Denote $L:=T'\cap T''$ be the common vertices of the
two unlabelled vertex sets, so $l=|L|$ ($0\leq l\leq k$) is the number
of common vertices in the unlabelled $k$ vertices. Denote $M'
:=T'\backslash L$, $M'' :=T''\backslash L$ be the sets of different
vertex sets between the two unlabelled vertex sets.
Then,
\begin{equation}\notag
\Delta (G',G'') = 
\left\{\begin{array}{ll}
                     \multirow{2}{*}{$2^l$}  & \textrm{if
                       $(S_1''\backslash M', S_2''\backslash M')$ is
                       matched by}\\
                     & \textrm{$(S_1'\backslash M'', S_2'\backslash
                       M'')$  or $(S_2'\backslash M'', S_1'\backslash
                       M'')$}\\  
                    0 & \textrm{otherwise}\\
                    \end{array}\right.
\end{equation}

\subsection{Counting Method for Edge Contraction Algorithm}
For EC (Alg. \ref{alg:edge-contraction}), after step $t$ ($t = 1,
\cdots, n-2$) there are $k = n-t$ ``super'' vertices (i.e.~contracted
ones).  It is straightforward to see that $|C(G')|=|C(G'')|=2^{k-1}-1$.

To count the intersection $\Delta (G',G'')$, suppose there are $l$ ($0\leq l\leq
k$) common super vertices in the unlabelled $k$ vertices. Remove the
$l$ common super vertices from each set, then there are $h = k-l$
distinct super vertices in each set, denote them by
$P:=\{\mathbf{p}_1,\mathbf{p}_2,\cdots,\mathbf{p}_h\}$,
$Q:=\{\mathbf{q}_1,\mathbf{q}_2,\cdots,\mathbf{q}_h\}$,
respectively. Notice that $\mathbf{p}_1\cup\mathbf{p}_2 \cup\cdots
\cup \mathbf{p}_h = \mathbf{q}_1 \cup\mathbf{q}_2\cup \cdots
\cup\mathbf{q}_h$, so after some contractions in both $P$ and $Q$,
there must be some common super vertices between $P$ and $Q$. Assume
the maximum number of common super vertices after all possible
contractions is $c^*$, then it holds 
\begin{equation}\label{}
  \Delta (G',G'')=2^{c^*+l-1}-1\;.
\end{equation}
To compute $c^*$, we propose a polynomial time algorithm
(Alg. \ref{alg:max-common}) with a theoretical guarantee in Theorem
\ref{theo:ec} (for the proof see Supplement \ref{app:proof-EC}). The
algorithm finds the maximal number of common super vertices after all
possible contractions, that is used to count $\Delta (G',G'')$ for EC.
\begin{theorem}
\label{theo:ec}
Given two distinct super vertex sets
$P:=\{\mathbf{p}_1,\mathbf{p}_2,\cdots,\mathbf{p}_h\}$,
$Q:=\{\mathbf{q}_1,\mathbf{q}_2,\cdots,\mathbf{q}_h\}$ (any 2 super
vertices inside $P$ or $Q$ do not intersect, and there is no common
super vertex between $P$ and $Q$), such that
$\mathbf{p}_1\cup\mathbf{p}_2 \cup\cdots \cup \mathbf{p}_h =
\mathbf{q}_1 \cup\mathbf{q}_2\cup \cdots \cup\mathbf{q}_h$,
Alg. \ref{alg:max-common} returns the maximum number of common super
vertices between $P$ and $Q$ after all possible contractions.
\end{theorem}
\begin{algorithm}
\begin{small}
  \caption{Common Super Vertex Counting
}\label{alg:max-common}
\KwIn{Two distinct super vertex sets $P$, $Q$ }
\KwOut{Maximum number of common super vertices after all possible contractions}
{$c:=0$\;}
\While{$P\neq \emptyset$}{
    {Randomly pick $\mathbf{p}_i\in P$\;}
    {Find $\mathbf{q}_j\in Q$ s.t. $\mathbf{p}_i\cap \mathbf{q}_j\neq \emptyset$\;} 
    \If{$\mathbf{q}_j\backslash\mathbf{p}_i\neq\emptyset$}{
        {For $\mathbf{p}_i$, find $\mathbf{p}_{i'}\in P\backslash \{\mathbf{p}_i\}$ s.t. $\mathbf{p}_{i'}\cap (\mathbf{q}_j\backslash\mathbf{p}_i)\neq\emptyset$\;} 
        {$\mathbf{p_{ii'}}:=\mathbf{p}_i\cup \mathbf{p}_{i'}$, $P:=P\cup\{\mathbf{p_{ii'}}\}\backslash \{\mathbf{p}_i,  \mathbf{p}_{i'}\}$ 
        }\;         
    }
    \If{$\mathbf{p}_i\backslash\mathbf{q}_j\neq\emptyset$}{
        {For $\mathbf{q}_j$, find $\mathbf{q}_{j'}\in Q\backslash\{\mathbf{q}_j\}$ s.t. $\mathbf{q}_{j'}\cap (\mathbf{p}_i\backslash\mathbf{q}_j)\neq\emptyset$\;}
        {$\mathbf{q_{jj'}}:=\mathbf{q}_j\cup \mathbf{q}_{j'}$, $Q:=Q\cup\{\mathbf{q_{jj'}}\}\backslash \{\mathbf{q}_j,  \mathbf{q}_{j'}\}$ 
        }\;    
    }
    \If{$\mathbf{p_{ii'}}==\mathbf{q_{jj'}}$}{
        {Remove $\mathbf{p_{ii'}}$, $\mathbf{q_{jj'}}$ from $P$, $Q$, respectively\;}
        {$c:=c+1$\;}  
    }
}
\Return{$c$}
\end{small}
\end{algorithm}

\section{Experiments}

We  conducted experiments on two exemplary models: the edge
reversal model and the Gaussian edge weights model. Each model
involves the master graph $G$ and a noise type used to generate the
two noisy instances $G^\prime$ and $G^{\prime\prime}$. The width of
the instance distribution is controlled by the strength of the noise
model. These models provide the setting to investigate the algorithmic
behavior.

\setkeys{Gin}{width=0.5\textwidth, height=0.35\textwidth}
\begin{figure*}[htbp]
\begin{center}
\subfloat[Edge Reversal Model, $n$: 100]{
\includegraphics{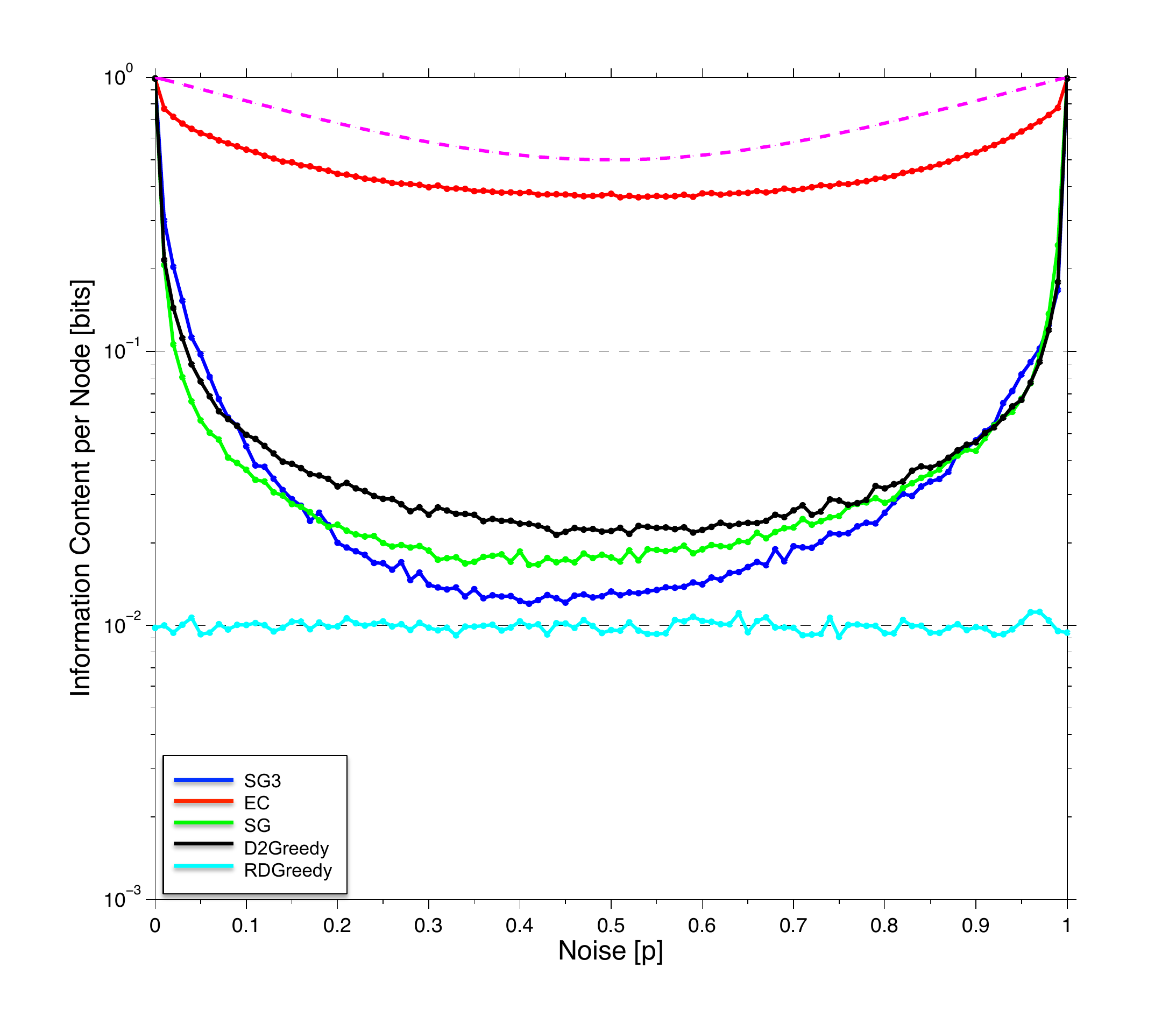}
}
\subfloat[Gaussian Edge Weights Model, $n$: 100]{
\includegraphics{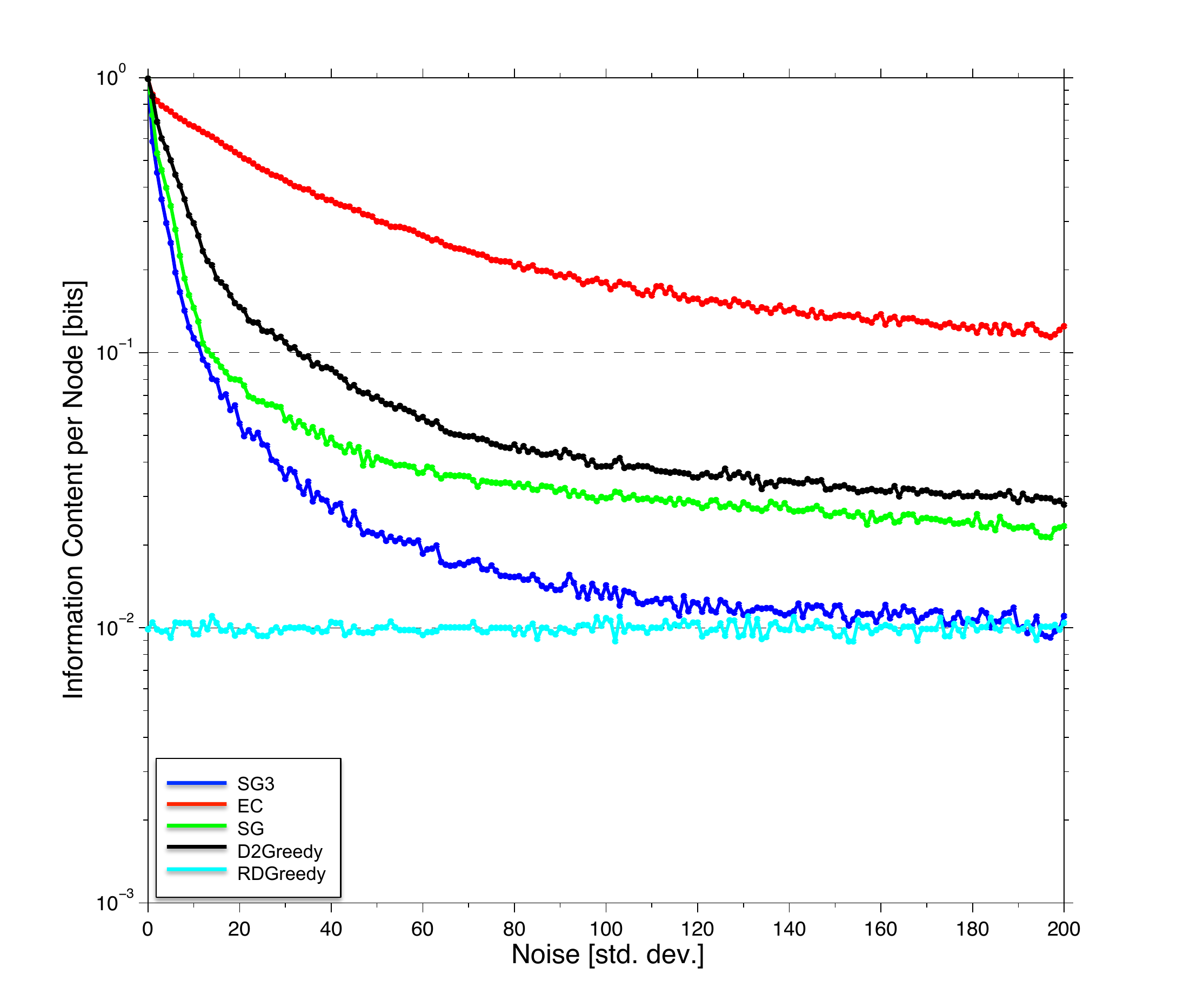}}
\end{center}
\caption{Information Content per Node}
\label{fig:ic}
\end{figure*}
\subsection{Experimental Setting}

\textbf{Edge Reversal Model}: 
%
To obtain the master graph, we generate a balanced bipartite graph
$G_b$ with disjoint vertex sets $S_1$, $S_2$.  Then we assign
uniformly distributed weights in $[0, \frac{8}{n^2}]$ to all edges
inside $S_1$ or $S_2$ and we assign uniformly distributed weights in
$[1- \frac{8}{n^2}, 1]$ to all edges between $S_1$ and $S_2$, thus
generating graph $G'_b$. Then randomly flip edges in $G'_b$ to
generate the master graph $G$. Here, flip edge
$e_{ij}$ means changing its weight $w_{ij}$ to $1-w_{ij}$ with
probability $p_m$, and $(flip\; e_{ij} ) \sim Ber
(p_m)$; $p_m = 0.2$ is used to generate the master graph $G$.
Noisy graphs $G'$, $G''$ are generated by flipping the edges in $G$
with probability $p$, ($(flip\; e_{ij} ) \sim Ber (p)$).

\textbf{Gaussian Edge Weights Model}: 
The master graph $G$ is generated with Gaussian distributed edge
weights $w_{ij} \sim N (\mu, \sigma_m^2)$, $\mu = 600, \sigma_m = 50$,
negative edges are set to be $\mu$.  {Noisy graphs $G'$, $G''$} are
obtained by adding Gaussian distributed noise $n_{ij} \sim N (0,
\sigma^2)$, negative noisy edges are set to be 0.

For both noise models, we conducted 1000 experiments on 
i.i.d. generated noisy graphs $G'$ and $G''$, and then we aggregate the
results to estimate the expectation in Eq.  \ref{eq:ic}.

\subsection{Results}

We plot the information content and stepwise information \textit{per
  node} in Fig. ~\ref{fig:ic} and ~\ref{fig:stepwise-info},
respectively. For the edge reversal model, we also investigate the
number of equal edge pairs between $G'$ and $G''$: $d = 0, \cdots, m$
($m$ is the total edge number), $d$ measures the consistency of the
two noisy instances.  The expected fraction of equal edge pairs is
$\mathbb{E} d = p^2 + (1 - p)^2$, and it is plotted as the dashed
magenta line in Fig. ~\ref{fig:ic}(a).


\subsection{Analysis}

Before discussing these results, let us revisit the stepwise
information and information content.  From the counting methods in
Section \ref{sec:counting}, we derive the analytical form of $|\C|$,
$|C_t^{\mathscr{A}}(G')|$ and $ |C_t^{\mathscr{A}}(G'')|$ (e.g., 
$\mathscr{A}=SG3$), an we insert these values into the definition of
stepwise information,
\begin{flalign}\label{eq:stepwise-sg3}
  &I_t^{\mathscr{A}} = \mathbb{E} \log \Bigl(|\C|
  \frac{\Delta_t^{\mathscr{A}}(G',G'')}{ |C_t^{\mathscr{A}}(G')|  |C_t^{\mathscr{A}}(G'')|} \Bigr)\\
  \notag & = \mathbb{E} (\log(|\C| \Delta_t^{\mathscr{A}}(G',G'')) -
  \log(|C_t^{\mathscr{A}}(G')|  |C_t^{\mathscr{A}}(G'')|))\\ 
  \notag & = \mathbb{E} \log\Delta_t^{\mathscr{A}}(G',G'') + 2t +
  \log(2^{n-1} - 1) -2(n - 1)
\end{flalign}
\setkeys{Gin}{width=0.5\textwidth, height=0.35\textwidth}
\begin{figure*}[htbp]
\begin{center}
\subfloat[Edge Reversal Model, $n$: 100, $p = 0.65$]{
\includegraphics{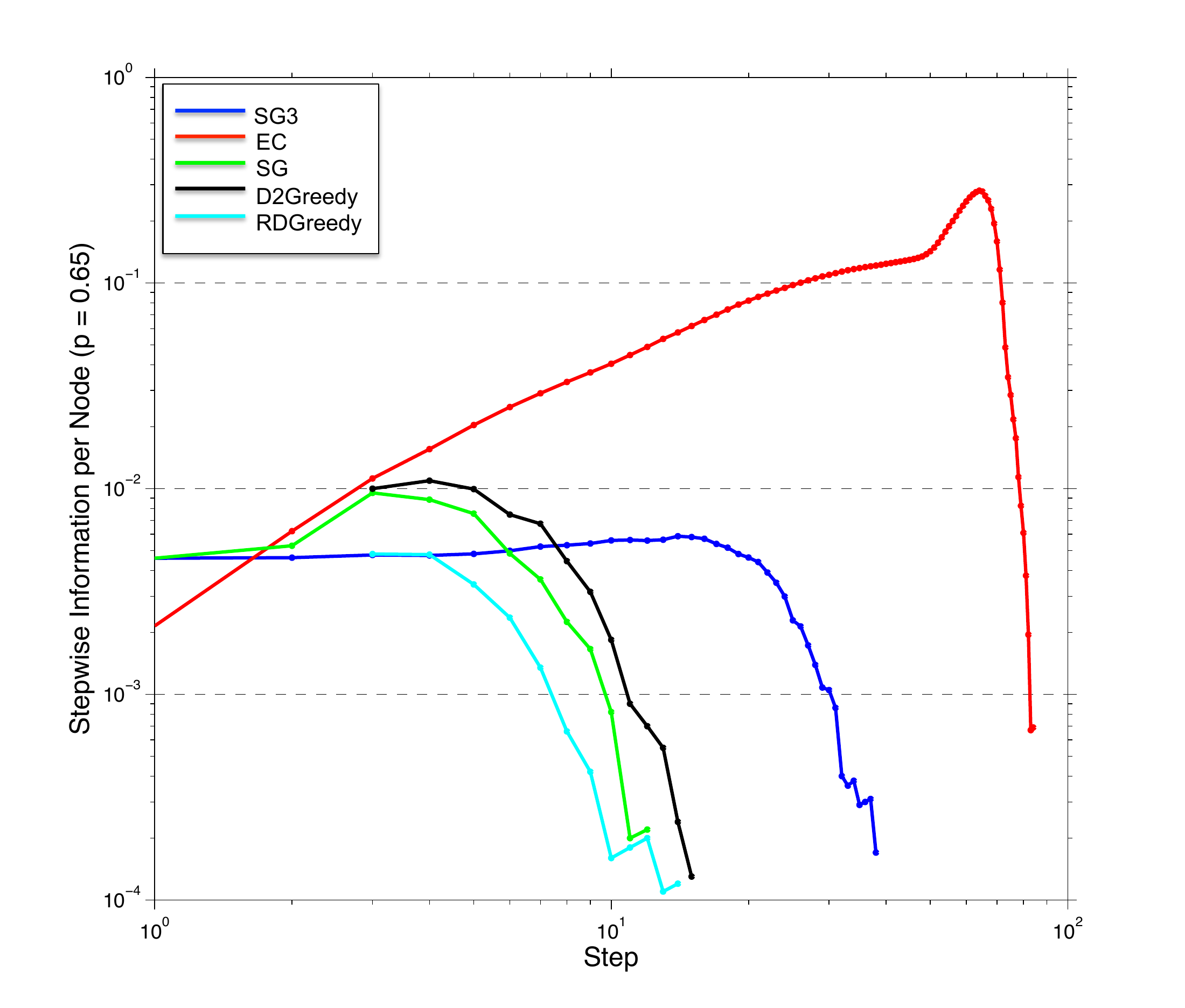}}
\subfloat[Gaussian Edge Weights Model, $n$: 100,  $\sigma=125$ ]{
\includegraphics{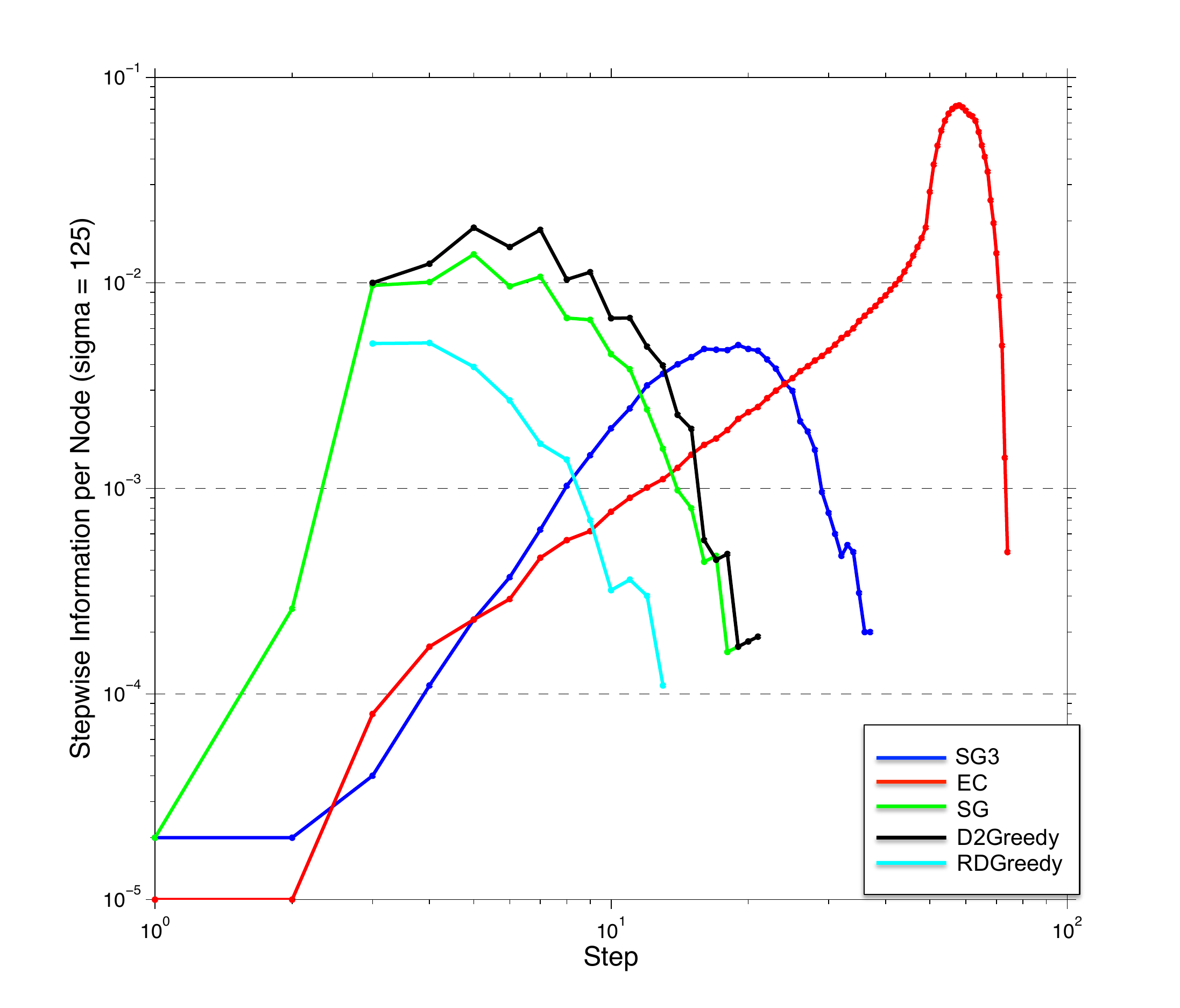}}
\end{center}
\caption{Stepwise Information per Node}
\label{fig:stepwise-info}
\end{figure*}
The information content is computed as the maximum stepwise
information $I^{\mathscr{A}} := \max_t I_t^{\mathscr{A}}$.  Notice
that $\log\Delta_t^{\mathscr{A}}(G',G'')$ measures the ability of
$\mathscr{A}$ to find common solutions for the two noisy instances
$G', G''$, given the underlying input graph $G$.

Our results support the following observations and analysis:

\sym All investigated algorithms reach the maximum information content
in the noise free limit ($G' = G''$), i.e., for $p=0, 1$ in the edge
reversal model and for $\sigma=0$ in the Gaussian edge weights
model. 
In this circumstance, 
$\mathbb{E} \log\Delta_t^{\mathscr{A}}(G',G'') =
\log|C_t^{\mathscr{A}}(G')| = n - t -1$, so $I_t^{\mathscr{A}} = t +
\log(2^{n-1} - 1) -(n - 1)$, and the information content reaches
its maximum $\log(2^{n-1} - 1)$ at the final step $t = n-1$.

\sym Fig.  \ref{fig:ic}(a) demonstrates that the information content
qualitatively agrees with the consistency between two noisy instances
(the dashed magenta line), which
 reflects that $\log\Delta_t^{\mathscr{A}}(G',G'')$ is affected by  the
 noisy instances.

\sym Stepwise information (Fig. ~\ref{fig:stepwise-info}) of the
algorithms increase initially, but after reaching the optimal
step $t^*$ (the step with highest information), it decreases and
finally vanishes.

\sym For the greedy heuristics, backward greedy is more informative
than double greedy under both models.  EC (backward greedy) achieves
the highest information content. We explain this behavior by delayed
decision making of the backward greedy edge contraction.  With high
probability it preserves consistent solutions by contracting low
weight edges that have a low probability to be included in the
cut. The same phenomena arises for the reverse-delete algorithm to
calculate the minimum spanning tree of a graph (see
\cite{informative-mst}).

\sym The information content of the four double greedy algorithms
achieve different rank orders for the two models.  SG3 is inferior to
other double greedy algorithms under Gaussian edge weights model, but
this only happens when $p\in [0.2, 0.87]$ for the edge reversal model.
This results from that information content of one specific algorithm is affected by 
both the input master graph $G$ and the noisy instances $G', G''$, 
which are completely different under the two models.

\sym Different greedy techniques cast different influences on the
information content.  The four double greedy algorithms differ by the
techniques they use (Table ~\ref{tab-alg-summarization}). (1) The
randomization technique makes RDGreedy very fragile w.r.t. information
content, though it
%
improves the worst-case approximation guarantee
for the general USM problem (\cite{buchbinder2012tight}).  RDGreedy
labels each vertex with a probability proportional to the objective
gain, this randomization makes the consistency between
$|C_t^{\mathscr{A}}(G')|$ and $|C_t^{\mathscr{A}}(G'')|$ very weak,
resulting in small approximation set intersection
$\log\Delta_t^{\mathscr{A}}(G',G'')$.  (2) The initializing strategy
for the first 2 vertices as used in SG decreases the information
content (SG is outperformed by D2Greedy under both models) due to
early decision making. (3) The situation is similar for the sorting
techniquey used in SG3 under Gaussian edge weights model, it is
outperformed by both SG and D2Greedy.  But for the edge reversal
model, this observation only holds when $p\in [0.2, 0.87]$.

\sym SG and D2Greedy behave very similar under both models, 
which is caused by an equivalent processing sequence apart from
initializing of the first two vertices (proved in Supplement
\ref{app:equivalence-SG-D2Greedy}).
%
%
%

%
%
%
%
%
%
%
%

\section{Discussion and Conclusion}

This work advocates an information theoretically guided average case
analysis of the generalization ability of greedy \MAXCUT\ algorithms.
We have contributed to the foundation of approximation set coding by
presenting provably correct methods to \textit{exactly} compute the
cardinality of approximation sets.  The counting algorithms for
approximate solutions enable us to explore the information content of
greedy \MAXCUT\ algorithms.  Based on the observations and analysis,
we propose the following conjecture:

\sym \textsl{Different greedy heuristics (backward, double) and different
  processing techniques (sorting, randomization, intilization) sensitively influence the information content. The
  backward greedy with its delayed decision making
  consistently outperforms the double greedy strategies for different
  noise models and noise levels. 
}

%
%
%

Since {{EC}} demonstrated to achieve the highest robustness, it is
valuable to develop the corresponding algorithm for the general USM.

In this work ASC has been employed as a descriptive tool to compare
algorithms. We could also use the method for algorithm design. A
meta-algorithm modifies the algorithmic steps of a \MAXCUT\ procedure
and measures the resulting change in information content. Beneficial
changes are accepted and detrimental changes are rejected. It is also
imaginable that design principles like delayed decision making are
systematically identified and then combined to improve the
informativeness of novel algorithms. 


\section*{Acknowledgment}
\begin{small}
  This work was partially supported by SNF Grant \# 200021 138117.  The authors would
  like to thank Andreas Krause, Mat\'us Mihal\'ak and Peter Widmayer for 
  valuable discussions.
\end{small}

\bibliographystyle{IEEEtran} 
\bibliography{ITW15}

\clearpage

\section{Supplementary Material}

\subsection{Details of Double Greedy Algorithms}\label{sup:alg}

\begin{algorithm}
\begin{small}
\caption{\textbf{SG} (\cite{sahni1976p})}\label{alg:sg}
\KwIn{A complete graph $G=(V, E)$ with nonnegative edge weights $w_{ij}, \forall i,j\in V, i\neq j$}
\KwOut{A disjoint cut and the cut value}
{Pick the maximum weighted edge $(x,y)$\;}
{$S_1:=\{x\}$,$S_2:=\{y\}$,  $cut(S_1, S_2):=w_{xy}$\;}
\For{$i = 1:n-2$}{
    {If $w(i, S_1)>w(i, S_2)$, then add $i$ to $S_2$, else add it to $S_1$\tcp*[r]{$w(i, S_k):= \sum_{j\in S_k}w_{ij}, k=1,2$}} 
    {$cut(S_1, S_2):=cut(S_1, S_2)+ \max\{w(i, S_1), w(i, S_2)\}$\;}
}
\Return{$S_1$, $S_2$, and $cut(S_1, S_2)$}
\end{small}
\end{algorithm}

\begin{algorithm}[htbp]
\begin{small}
\caption{\textbf{RDGreedy} (\cite{buchbinder2012tight})}\label{alg:r-usm-4-max-cut}
\KwIn{A complete graph $G=(V, E)$ with nonnegative edge weights $w_{ij}, \forall i,j\in V, i\neq j$}
\KwOut{A disjoint cut  and the cut value }
{$S^0 :=\emptyset$, $T^0 := V$\;}
\For{$i=1$ to $n$}{
    {$a_i := f(S^{i-1} \cup \set{v_i})-f(S^{i-1})$\;}
    {$b_i := f(T^{i-1} \backslash \set{v_i})-f(T^{i-1})$\;}
    {$a_i':=\max\{a_i, 0\}$, $b_i':=\max\{b_i, 0\}$\;}
    {\textbf{With probability} $\frac{a_i'}{a_i'+b_i'}$ \textbf{do}: $S^i :=S^{i-1} \cup \set{v_i}$, $T^{i}:=T^{i-1}$ \tcp{If $a_i'=b_i'=0$, assume $\frac{a_i'}{a_i'+b_i'}=1$}}
    {\textbf{Else} (with the compliment probability $\frac{b_i'}{a_i'+b_i'}$) \textbf{do}:}
    {$S^i :=S^{i-1}$, $T^{i}:=T^{i-1} \backslash \set{v_i}$\;}
}
\Return{2 subsets: $S^{n}$, $V \backslash S^{n}$, and $cut(S^{n}, V \backslash S^{n})$}
\end{small}
\end{algorithm}
\begin{algorithm}
\begin{small}
\caption{\textbf{SG3}}\label{alg:sg3}
\KwIn{A complete graph $G=(V, E)$ with nonnegative edge weights $w_{ij}, \forall i,j\in V, i\neq j$}
\KwOut{A disjoint cut $S_1, S_2$ and the cut value $cut(S_1, S_2)$}
{Pick the maximum weighted edge $(x,y)$\;}
{$S_1:=\{x\}$,$S_2:=\{y\}$,$V:=V\backslash\{x,y\}$, $cut(S_1, S_2):=w_{xy}$\;}
\For{$i=1:n-2$}{
    \For{$j\in V$}{
        {$score(j):=|w(j, S_1)-w(j, S_2)|$ \tcp*[r]{$w(j, S_k):= \sum_{j'\in S_k}w_{jj'}, k=1,2$}} 
    }
    {Choose the vertex $j^*$ with the maximum score\;}
    {If $w(j^*, S_1)>w(j^*, S_2)$, then add $j^*$ to $S_2$, else add it to $S_1$\;}
    {$V:=V\backslash\{j^*\}$\;}
    {$cut(S_1, S_2):=cut(S_1, S_2) + \max\{w(j^*, S_1), w(j^*, S_2)\}$\;}
}
\Return{$S_1$, $S_2$, and $cut(S_1, S_2)$}
\end{small}
\end{algorithm}

\subsection{Equivalence Between Labelling Criterions of SG and D2Greedy}
\label{app:equivalence-SG-D2Greedy}

\textbf{Claim}: \quad
Except for processing the first 2 vertices, D2Greedy and SG conduct the same labelling strategy for each
vertices.
\begin{proof}
To verify this, assume in the beginning of a  certain step $i$, the solution set pair  of SG is $(S_1, S_2)$, of
D2Greedy is $(S, T)$ (for simplicity  omit the step index here).  

Note that the relationship between solution sets of SG and D2Greedy is: $S_1 \leftrightarrow S$ and 
$S_2 \leftrightarrow (V \backslash T)$. 

For SG, the labelling criterion for vertex $i$ is: 
\begin{equation}
	w(i, S_2) - w(i, S_1) = \sum_{i,  j\in S_2} w_{ij} -  \sum_{i,  j\in S_1} w_{ij} 
\end{equation}

For D2Greedy, the labelling criterion for vertex $i$ is: 

\begin{flalign}
\notag	a_i - b_i & = [f(S \cup \set{v_i}) - f(S) ] - [f(T \backslash \set{v_i}) -f(T) ] \\
\notag				& = \left( \sum_{i\in S \cup \set{v_i}, j\in V \backslash S \backslash \set{v_i}} w_{ij} -  \sum_{i\in S, j\in V \backslash S} w_{ij} \right) -  \\  
				& {\phantom = } \left( \sum_{i\in T \backslash \set{v_i}, j\in V \backslash T \cup \set{v_i}} w_{ij} -  \sum_{i\in T, j\in V \backslash T} w_{ij} \right)  \\
\notag				& = \left( \sum_{i, j\in V \backslash S \backslash \set{v_i}} w_{ij} -  \sum_{i\in S, j = i} w_{ij} \right) -  \\  
				& {\phantom = } \left( \sum_{i\in T \backslash \set{v_i}, j = i} w_{ij} -  \sum_{i, j\in V \backslash T} w_{ij}\right)  \\	
\notag				& = \left( \sum_{i, j\in (V \backslash T) \cup (T \backslash S \backslash \set{v_i})} w_{ij} -  \sum_{i, j\in S} w_{ij}\right) -  \\  
				& {\phantom = } \left( \sum_{i, j \in (S) \cup (T \backslash S \backslash \set{v_i})} w_{ij} -  \sum_{i, j\in V \backslash T} w_{ij}\right)  \\
\notag			& = 2\left(\sum_{i, j\in V \backslash T} w_{ij} - \sum_{i, j\in S} w_{ij} \right)\\
\label{eq:relation}   & =  2\left(\sum_{i, j\in S_2} w_{ij} - \sum_{i, j\in S_1} w_{ij}  \right)\\
\notag 				& = 2[w(i, S_2) - w(i, S_1)]				   
\end{flalign}
\noindent where Eq. \ref{eq:relation} comes from the relationship between solution sets of SG and D2Greedy.

So the labelling criterion for SG and D2Greedy is equivalent with each other. 
\end{proof}

\subsection{Counting Methods for  Double Greedy Algorithms}
\label{complement:counting}

\textbf{D2Greedy}: summarized in  Alg. \ref{alg:d-usm-4-max-cut}, we have proved that it has the same labelling criterion
with SG, the  relationship between solution sets of SG and D2Greedy is: $S_1 \leftrightarrow S$ and 
$S_2 \leftrightarrow (V \backslash T)$, we will use $S_1$ and $S_2$ in the description of its 
counting methods. 

In  step $t$ ($t = 1, \cdots, n$) there are $k = n-t$ unlabelled vertices, it is not difficult to know that  the number of possible solutions for each instance is

\begin{equation}\notag
|C(G')| = |C(G'')| = \left\{\begin{array}{ll}
                     2^{k}  & \textrm{if $S_1 \neq \emptyset$ and $S_2 \neq \emptyset$}\\
                     2^{k}-1  & \textrm{otherwise}\\
                    \end{array}\right.
\end{equation}

To count the intersection set (i.e. $|C(G')\cap C(G'')|$),
assume the solution sets of $G'$ is $(S_1', S_2')$, the solution sets of $G''$ is $(S_1'', S_2'')$, so the unlabelled vertex sets are $T'=V\backslash S_1'\backslash S_2'$, $T''=V\backslash S_1''\backslash S_2''$,
respectively.
Denote $L:=T'\cap T''$ be the common vertices of the two unlabelled vertex sets,
so $l=|L|$ ($0\leq l\leq k$) is the number of common vertices in the unlabelled
$k$ vertices. Denote $M' :=T'\backslash L$, $M'' :=T''\backslash L$
be the sets of different vertex sets between the two unlabelled vertex sets.
Then,

\begin{enumerate}

\item if  $(S_1'\backslash M'', S_2'\backslash M'')$ or $(S_2'\backslash M'', S_1'\backslash M'')$ matches 
$(S_1''\backslash M', S_2''\backslash M')$.

Assume w.l.o.g. that $(S_1'\backslash M'', S_2'\backslash M'')$  matches 
$(S_1''\backslash M', S_2''\backslash M')$:

  \begin{equation}\notag
  \begin{split}
& |C(G^{'})\cap C(G^{''})| = \\
& \left\{\begin{array}{ll}
                     2^{l}  & \textrm{if $S_1' \cup S_1'' \neq \emptyset$ and $S_2' \cup S_2'' \neq \emptyset$}\\
                     2^{l}-1  & \textrm{otherwise}\\
                    \end{array}\right.
\end{split}
\end{equation}

\item  otherwise, $|C(G^{'})\cap C(G^{''})|$ =  0 
\end{enumerate}

\textbf{SG3}:  presented in Section  \ref{sec:counting-sg3}.

\textbf{SG}: summarized in Alg. \ref{alg:sg}, the methods to count its approximation sets is 
the same as that of SG3.

\textbf{RDGreedy}: summarized in Alg. \ref{alg:r-usm-4-max-cut}, the methods to count its approximation sets is 
the same as that of D2Greedy.

\subsection{Proof of the Correctness of Method to Count $|C(G')\cap C(G'')|$ of SG3}
\label{app:proof-SG3}

\begin{proof}

First of all, notice that $M'$ must be included in $S_1''\cup S_2''$  and $M''$ must be included in $S_1'\cup S_2'$, because
$M'$ has no intersection with $M''$, and we know that $S_1''\cup S_2''\cup M''=S_1'\cup S_2'\cup M'$. After removing $M'$ from $S_1''\cup S_2''$, and  $M''$ from $S_1'\cup S_2'$, the vertices in the
pairs, $(S_1'\backslash M'', S_2'\backslash M'')$ and
  $(S_1''\backslash M', S_2''\backslash M')$, can not be changed by distributing any other unlabelled vertices
 , so if they can not match with each other, there will be no common solutions.

  If they can match, in the following, there is only one way to distribute  $M'$ and $M''$
  to have common solutions. And the vertices in the common set $L=T'\cap T''$ can be distributed consistently
  in the two instances, so in this situation $|C(G')\cap C(G'')|=2^l$.
\end{proof}

\subsection{Proof of Theorem \ref{theo:ec}}
\label{app:proof-EC}
\begin{proof}
First of all, 
We will prove the following claim, then use the claim to prove Theorem \ref{theo:ec}. 

\textbf{Claim}: \quad
In each step $t$ ($t = 0, \cdots, n-2$), the following conditions hold:
\begin{enumerate}

\item The remained super vertices in $P, Q$ are distinct with each other, that means 
any 2 super vertices inside $P$ or  $Q$ do not have intersection, and 
there are no common super 
vertex between $P$ and $Q$.

\item The common super vertex removed from $P, Q$, i.e., $\mathbf{p_{ii'}}=\mathbf{q_{jj'}}$,
is the smallest common super vertex containing $\mathbf{p}_i$ or $\mathbf{p}_{i'}$ (respectively, $\mathbf{q}_j$ or $\mathbf{q}_{j'}$)

\item The common super vertex removed from $P, Q$, i.e., $\mathbf{p_{ii'}}=\mathbf{q_{jj'}}$,
are ``unique'' (i.e., there does not exist $\mathbf{p_{ii''}}=\mathbf{q_{jj''}}$, such that
$\mathbf{p_{ii''}} \neq \mathbf{p_{ii'}}$). That means, there is only one possible way to construct the 
removed common super vertex. 

\end{enumerate}

We will use inductive assumption  to prove the {claim}.
First of all, in the beginning (step 0), the conditions hold. Assume the conditions hold in step $t$.
In step $t+1$, there are 2 possible situations:

\begin{itemize}
\item There are no common super vertex removed.

Condition 1 holds because the contracted super vertices pair do not equal. Condition 2, 3 hold as
well because there are no contracted super vertices removed.

\item There are common super vertex removed.

Condition 1 holds because the only common super vertices pair have been removed from $P, Q$, respectively.\\

To prove condition 2, notice that the smaller vertices for
$\mathbf{p_{ii'}}$ are $\mathbf{p_{ii'}} \backslash
\mathbf{p_{i}}=\mathbf{p_{i'}}$ and $\mathbf{p_{ii'}} \backslash
\mathbf{p_{i}'}=\mathbf{p_{i}}$, respectively, for $\mathbf{q_{jj'}}$
are $\mathbf{q_{jj'}} \backslash \mathbf{q_{j}}=\mathbf{q_{j'}}$ and
$\mathbf{q_{jj'}} \backslash \mathbf{q_{j}'}=\mathbf{q_{j}}$,
according to Condition 1, they can not be common
super vertices, so there are no smaller common super vertices.\\

To prove condition 3, assume there exists
$\mathbf{p_{ii''}}=\mathbf{q_{jj''}}$, such that $\mathbf{p_{ii''}}
\neq \mathbf{p_{ii'}}$ (respectively, $\mathbf{q_{jj''}} \neq
\mathbf{q_{jj'}}$), so $\mathbf{p_{i''}}\neq \mathbf{p_{i'}}$
($\mathbf{p_{j''}}\neq \mathbf{p_{j'}}$). From
Alg. \ref{alg:max-common} we know that $\mathbf{p_{i}} \cup
\mathbf{p_{i''}}=\mathbf{p_{ii''}} \supseteq \mathbf{q_{j}} \backslash
\mathbf{p_{i}}$ and $\mathbf{p_{i}}
\cup\mathbf{p_{i'}}=\mathbf{p_{ii'}} \supseteq \mathbf{q_{j}}
\backslash \mathbf{p_{i}}$ (respectively, $\mathbf{q_{j}} \cup
\mathbf{q_{j''}}=\mathbf{q_{jj''}} \supseteq \mathbf{p_{i}} \backslash
\mathbf{q_{j}}$ and $\mathbf{q_{j}} \cup
\mathbf{q_{j'}}=\mathbf{q_{jj'}} \supseteq \mathbf{p_{i}} \backslash
\mathbf{q_{j}}$), so that $\mathbf{p_{i''}} \supseteq \mathbf{q_{j}}
\backslash \mathbf{p_{i}}$ and $\mathbf{p_{i'}}\supseteq
\mathbf{q_{j}} \backslash \mathbf{p_{i}}$ (respectively,
$\mathbf{q_{j''}} \supseteq \mathbf{p_{i}} \backslash \mathbf{q_{j}}$
and $\mathbf{q_{j'}}\supseteq \mathbf{p_{i}} \backslash
\mathbf{q_{j}}$), that contradicts the known truth that
$\mathbf{p_{i'}}$ and $\mathbf{p_{i''}}$ (respectively,
$\mathbf{q_{j'}}$ and $\mathbf{q_{j''}}$) must be totally different
with each other (from Condition 1).

\end{itemize}

Then we use the Claim to prove that the $c$ returned by Alg. \ref{alg:max-common} is exactly the maximum number of common super vertices after all possible
contractions.
Because the 3 conditions hold for each step, we know that finally all the common super vertices
are removed out from $P$ and $Q$.
From Condition 2 we know that all the removed common super vertices are the smallest ones, from Condition
3 we get that there is not a second way to construct the common super vertices, 
 so the resulted $c$ is the maximum number of common super vertices
after all possible contractions.

\end{proof}

\end{document}